\documentclass[a4paper]{article}

\usepackage{INTERSPEECH2016}

\usepackage{graphicx}
\usepackage{amssymb,amsmath,bm}
\usepackage{amsopn}
\usepackage{textcomp}
\usepackage{multirow}
\usepackage{hhline}
\usepackage{makecell}
\usepackage{tikz}
\usepackage{verbatim}
\usepackage{fixltx2e}
\usepackage{textcomp}
\usepackage{subfig}
\usepackage{array}

\newcommand{\thickhline}{%
    \noalign {\ifnum 0=`}\fi \hrule height 1pt
    \futurelet \reserved@a \@xhline
}
\newcolumntype{"}{@{\hskip\tabcolsep\vrule width 1pt\hskip\tabcolsep}}
\makeatother

\DeclareMathOperator*{\argmax}{arg\,max}

\ninept

\title{Joint Sound Source Separation and Speaker Recognition}

\makeatletter
\def\name#1{\gdef\@name{#1\\}}
\makeatother \name{{\em Jeroen Zegers, Hugo Van hamme}}

\address{KU Leuven, Dept. ESAT, Belgium \\
}

\begin{document}

  \maketitle
  \begin{abstract}
Non-negative Matrix Factorization (NMF) has already been applied to learn speaker characterizations from single or non-simultaneous speech for speaker recognition applications. It is also known for its good performance in (blind) source separation for simultaneous speech. This paper explains how NMF can be used to jointly solve the two problems in a multichannel speaker recognizer for simultaneous speech. It is shown how state-of-the-art multichannel NMF for blind source separation can be easily extended to incorporate speaker recognition. Experiments on the CHiME corpus show that this method outperforms the sequential approach of first applying source separation, followed by speaker recognition that uses state-of-the-art i-vector techniques.

  \end{abstract}

  \section{Introduction}
  

  Non-negative Matrix Factorization (NMF) is a frequently used method to identify patterns in data. It was originally developed to recognize parts-based representation in images \cite{lee1999learning} and has since been used in various fields such as bioinformatics \cite{gao2005improving}, noise-robust automatic speech recognition \cite{gemmeke2011exemplar} and age and gender estimation \cite{bahari2011speaker}. Recently it has been used for speaker recognition (SR) tasks and obtained comparable results to state-of-the-art i-vector based approaches \cite{joder2012exploring,drgas2015speaker}. Moreover, NMF has been used in scenarios of source separation (SS) and music transcription \cite{cichocki2009nonnegative,virtanen2007monaural}. Multichannel extensions of NMF have been developed with applications to blind source separation \cite{sawada2013multichannel,ozerov2010multichannel}. These approaches combine spatial cues from phase differences between microphones and the segmentation strength from NMF without any prior knowledge of the sources.
  
State-of-the-art speaker recognition systems are made robust to speaker and inter-channel variability by determining a total variability space \cite{dehak2011front}. Also, attempts have been made at robustness to noise and reverberation \cite{zhao2014robust}. However, these models cannot cope with simultaneous speech, where multiple speakers are active at the same time. In such a case one has to resort to a solution where first source separation is applied, followed by a single speaker recognizer \cite{may2012binaural}. 
Overlapping speech occurs frequently in conversations. For example, in a telephone conversation, a meeting, a panel discussion or natural speech in general. Instead of solving the problems of source separation and speaker recognition sequentially, it is shown how NMF is inherently capable of solving these two problems jointly. 
In this paper synthetic convolutive mixtures are used to analyze the problem.

  This paper is organized as follows. Section \ref{NMFexplained} explains the basics of NMF and how it can be used in both SR and (multichannel) SS. Section \ref{JointApproach} shows how NMF can jointly solve these problems. Experiments are explained in section \ref{experiments} and a conclusion is given in section \ref{conclusion}.
  
  \section{Non-Negative Matrix Factorization}
	\label{NMFexplained}
\subsection{Principle}
	Non-negative matrix factorization is a factorization method that approximates a non-negative matrix $\mathbf{X} \in \mathbb{R}_+^{F \times N}$ using a non-negative dictionary matrix $\mathbf{T} \in \mathbb{R}_+^{F \times K}$ and a non-negative activation matrix $\mathbf{V} \in \mathbb{R}_+^{K \times N}$, such that $\mathbf{X} \approx \hat{\mathbf{X}} \triangleq \mathbf{TV}$. In our application $\mathbf{X}=\vert \tilde{\mathbf{X}} \vert^{.2}$ is a speech power spectrogram, with $\tilde{\mathbf{X}}$ the complex valued short time Fourier transform (STFT) of the audio signal, $|.|$ the absolute value and $^{.2}$ the element wise square. $\mathbf{X}$ is a matrix with $F$ frequency bins and $N$ time frames. NMF tries to capture the most frequent patterns of the speech in $K$ $F$-dimensional basis vectors that form a dictionary $\mathbf{T}$ for the speech. The matrix $\mathbf{V}$ contains the coefficients of the linear combination and thus indicates how the $k^{th}$ basis vector is activated in the $n^{th}$ time frame. Usually $K < min(F,N)$ such that NMF is a rank reduction operation. A discrepancy measure is chosen between the original $\mathbf{X}$ and the reconstruction $\hat{\mathbf{X}}$ and can be minimized by finding optimal dictionaries and activations. The Euclidian (EU) distance, the Kullback-Leibler (KL) divergence and the Itakura-Saito (IS) divergence are well known measures. In this paper the IS divergence will be used.
	\begin{equation}
	\label{ISdiv}
	d_{IS}(x_{fn},\hat{x}_{fn})= \frac{{x}_{fn}}{\hat{{x}}_{fn}}-log\left( \frac{{x}_{fn}}{\hat{{x}}_{fn}}\right)  - 1
	\end{equation}
To minimize this divergence, multiplicative iterative update formulas with convergence guarantees have been derived \cite{nakano2010convergence}.
 
	\begin{equation}
	\label{Wupdate}
t_{fk} \leftarrow t_{fk} \sqrt{\frac{\sum_n \frac{x_{fn}}{\hat{x}_{fn}}\frac{v_{kn}}{\hat{x}_{fn}}}{\sum_n \frac{v_{kn}}{\hat{x}_{fn}}}}
	\end{equation}
		\begin{equation}
	\label{Hupdate}
v_{kn} \leftarrow v_{kn} \sqrt{\frac{\sum_f \frac{x_{fn}}{\hat{x}_{fn}}\frac{t_{fk}}{\hat{x}_{fn}}}{\sum_f \frac{t_{fk}}{\hat{x}_{fn}}}}
	\end{equation}
	where the sub-indices refer to the corresponding element in the matrix. To avoid scaling ambiguities the columns of $\mathbf{T}$ are to be normalized: $t_{fk} \leftarrow t_{fk}/\sum_{f^*} t_{f^*k}$.



	
    \subsection{NMF in Speaker Recognition}
    \label{NMFforSR}
The use of NMF in SR applications for single speech is straightforward. 
In the training phase, training data of the $j^{th}$ target speaker $\mathbf{X}_{train}^j$  is factorized using equations \ref{Wupdate} and \ref{Hupdate}. The obtained dictionaries $\mathbf{T}^j$, for each of the $J$ target speakers, are assumed to be speaker dependent and are collected in the library $\mathbf{T}_{tot}=[\mathbf{T}^1, \mathbf{T}^2, \ldots, \mathbf{T}^J]$.

During testing the identity of a speaker $s$ has to be found in a previously unseen $\mathbf{X}_{test}^s$. NMF is applied with a fixed library $\mathbf{T}_{tot}$ and the activations $\mathbf{V}_{tot,test}^s$ are found iteratively using equation \ref{Hupdate}. The activation matrix quantitatively indicates the activation of each basis vector for each target speaker in each time frame. The combined activity of all basis vectors in a target speaker's dictionary is a measure of the activity of the target speaker in the test segment. It is possible to include Group Sparsity (GS-NMF) constraints on the activations $\mathbf{V}_{tot,test}^s$ to enforce solutions where it is unlikely that basis vectors from different target speakers are active at the same time frame \cite{lefevre2011itakura,hurmalainen2012group}. A simple way of estimating the speaker identity is to determine the target speaker for which the sum of the activations, over all its basis vectors and over all the time frames, is maximal. This way of classification can be seen as a per frame speaker probability estimation where the final estimation is a weighted average over all frames, giving more weight to frames with higher activation or more energy.
\begin{equation}
\label{SIDestH}
\hat{ID}_s= \argmax_j \sum_{k \in \kappa^j} \sum_n v_{tot,kn}^s
\end{equation}
where $\kappa^j$ are the indices of the basis vectors belonging to the dictionary of the $j^{th}$ target speaker.
It is possible to perform a more advanced classification of the activations to a speaker identity by using, for example, support vector machines.



    \subsection{NMF in Source separation}
    \label{NMFforSS}
    Aside from SR applications, NMF has also shown good results in source separation problems. When the different speakers are learned on single speech training data, the procedure is very similar to that of SR. However, in SS, the test data $\mathbf{X}_{test}$ contains speech of multiple sources that speak simultaneously. The task is not to determine the speaker identity, but the original source signal of each speaker.

    After learning $\mathbf{T}_{tot}$ in the training phase, $\mathbf{V}_{tot,test}$ is calculated in the same way as in section \ref{NMFforSR}.
    Using Wiener filtering and the phase of the observations, the original source signal $\hat{y}_{fn}^s$ can be estimated \cite{virtanen2007monaural}.
\begin{equation}
\hat{y}_{fn}^s=\left(\frac{\sum_{k \in \kappa^s} t_{fk}v_{kn}}{\sum_{s^*} \sum_{k \in \kappa^{s^*}}t_{fk}v_{kn}}|\tilde{x}_{fn}|\right)arg(\tilde{x}_{fn})
\end{equation}
where $\kappa^s$ are the indices of the basis vectors belonging to the dictionary of the $s^{th}$ speaker and $arg(\tilde{x}_{fn})$ denotes the phase of $\tilde{x}_{fn}$.

In many situations, however, there is no possibility for supervised source separations.  
In blind source separation (BSS) no $\mathbf{X}_{train}^s$ is available to learn the library $\mathbf{T}_{tot}$. Instead the library will be created during the separation itself. Usually one resorts to multichannel techniques in such cases, where Time Direction Of Arrival (TDOA) techniques can be used to assist the source separation.
%
%
The mixing matrix $\mathbf{M} \in \mathbb{C}^{F \times I \times S}$ is assumed static and thus independent on $n$.
\begin{equation}
\label{mixture}
\tilde{x}_{i,fn}=\sum_s m_{is,f}y_{s,fn}
\end{equation}
where $I$ is the amount of microphones and $m_{is,f}$ indicates the frequency domain representation of the room impulse response (RIR) between the $s^{th}$ speaker and the $i^{th}$ microphone for the $f^{th}$ frequency bin. $\tilde{x}_{i}$ is the received STFT spectrogram in the $i^{th}$ microphone and $y_s$ is the STFT spectrogram of the original signal of the $s^{th}$ speaker. Because of the scaling ambiguity in equation \ref{mixture}, only the relative RIRs between the microphones can be estimated. The notation for the combined microphone signals is as follows.
\begin{equation}
\tilde{\mathbf{x}}_{fn}=[\tilde{x}_{1,fn},\tilde{x}_{2,fn}, \ldots, \tilde{x}_{I,fn}]
\end{equation}
Sawada et al. proposed a multichannel IS divergence \cite{sawada2013multichannel}.
\begin{align}
\label{multiISdiv}
\begin{split}
 D_{IS}(\mathbf{X},\lbrace\mathbf{T},\mathbf{V},\mathbf{H},\mathbf{Z}\rbrace)=\sum_f\sum_n d_{IS}(X_{fk},\hat{X}_{fk})
\\
 d_{IS}(X_{fk},\hat{X}_{fk})=tr(X_{fk}\hat{X}_{fk})-log det(X_{fk}\hat{X}_{fk})
\end{split}
\end{align}
where $tr(.)$ is the trace of a matrix, $logdet(.)$ is the natural logarithm of the determinant of a matrix, $X_{fn}=\tilde{\mathbf{x}}_{fn}\tilde{\mathbf{x}}_{fn}^H$ with $.^H$ the Hermitian transpose of a matrix and $\hat{X}_{fn}=\sum_k \left(\sum_s H_{fs} z_{sk}\right)t_{fk}v_{kn}$.
The same interpretations are given to $t_{fk}$ and $v_{kn}$ as in single channel NMF. 
$z_{sk}$ is a latent speaker-indicator that indicates the certainty that the $k^{th}$ basis vector belongs to the dictionary of the $s^{th}$ speaker under the constraints $z_{sk} \geq 0$ and $ \sum_s z_{sk} =1$. $H_{fs}$ is an $I \times I$ Hermitian positive semi-definite matrix with on its diagonals the power gain of the $s^{th}$ speaker at the $f^{th}$ frequency bin to each microphone. The off-diagonal elements include the phase differences between microphones and thus contain spatial information of the speaker. 
Multiplicative update formulas have been found in \cite[eq. 42-47]{sawada2013multichannel} that minimize the divergence in equation \ref{multiISdiv}.
The separated signals are then obtained through Wiener filtering.
\begin{equation}
\label{estSMultiNMF}
\hat{y}_{fn}^s=\left(\sum_k z_{sk}t_{fk}v_{kn}\right)H_{fs}\hat{X}^{-1}_{fn}\tilde{\mathbf{x}}_{fn}
\end{equation}

  \section{Joint approach}
  \label{JointApproach}
When performing speaker recognition in simultaneous speech scenarios, one could opt for a sequential approach. First apply blind source separation to obtain multiple, supposedly single speech, segments from simultaneous speech. Proceed as if those segments do not contain any crosstalk and apply single speaker recognition as explained in section \ref{NMFforSR}. However in this paper a joint approach is chosen, where speakers are characterized through dictionaries while separating the sources.
During training, source separation is performed as explained in section \ref{NMFforSS}. The $k^{th}$ basis vector is then assigned to the $s^{th}$ speaker for which $z_{sk}$ is maximal. The dictionaries are collected in the library $\mathbf{T_{tot}}$.
While testing, a similar source separation algorithm is used but $\mathbf{T_{tot}}$ remains fixed. Since every basis vector is contained in a dictionary, the meaning of the $\mathbf{Z}$ variable is changed. It now maps a complete dictionary $j$, and its corresponding target speaker identity, to a test speaker $s$. A new variable indicator $c_{jk}$ is introduced that assigns a basis vector $k$ to a dictionary $j$ if $c_{jk}=1$ under the constraints $c_{j^*k} \geq 0$ and $ \sum_{j^*} c_{j^*k} =1$. The variable $\hat{X}_{fn}$ is then reformulated as follows.
\begin{equation}
\hat{X}_{fn}=\sum_k \sum_j\sum_s H_{fs} z_{sj}c_{jk}t_{fk}v_{kn}
\end{equation}
It can be easily shown that the update formulas below then extend \cite[eq. 42-47]{sawada2013multichannel}. 
\begin{equation}
t_{fk} \leftarrow t_{fk} \sqrt{\frac{\sum_j c_{jk}\sum_s z_{sj} \sum_n v_{kn} tr(\hat{X}^{-1}_{fn}X_{fn}\hat{X}^{-1}_{fn}H_{fs})}{\sum_j c_{jk}\sum_s z_{sj} \sum_n v_{kn} tr(\hat{X}^{-1}_{fn}H_{fs})}}
\end{equation}
\begin{equation}
v_{kn} \leftarrow v_{kn} \sqrt{\frac{\sum_j c_{jk}\sum_s z_{sj} \sum_f t_{fk} tr(\hat{X}^{-1}_{fn}X_{fn}\hat{X}^{-1}_{fn}H_{fs})}{\sum_j c_{jk}\sum_s z_{sj} \sum_f t_{fk} tr(\hat{X}^{-1}_{fn}H_{fs})}}
\end{equation}
\begin{equation}
\label{zUpdtest}
z_{sj} \leftarrow z_{sj} \sqrt{\frac{\sum_k c_{jk}\sum_{f}\sum_{n} t_{fk}v_{kn} tr(\hat{X}^{-1}_{fn}X_{fn}\hat{X}^{-1}_{fn}H_{fs})}{\sum_k c_{jk}\sum_{f}\sum_{n} t_{fk}v_{kn} tr(\hat{X}^{-1}_{fn}H_{fs})}}
\end{equation}
\begin{equation}
\label{CUpdtest}
c_{jk} \leftarrow c_{jk} \sqrt{\frac{\sum_s z_{sj}\sum_{f}\sum_{n} t_{fk}v_{kn} tr(\hat{X}^{-1}_{fn}X_{fn}\hat{X}^{-1}_{fn}H_{fs})}{\sum_s z_{sj}\sum_{f}\sum_{n} t_{fk}v_{kn} tr(\hat{X}^{-1}_{fn}H_{fs})}}
\end{equation}
To update $H_{fs}$ an algebraic Riccati equation is solved. 
\begin{equation}
H_{fs}AH_{fs}=B
\end{equation}
\begin{equation}
A=\sum_j\sum_k c_{jk} z_{sj}t_{fk}\sum_n v_{kn}\hat{X}^{-1}_{fn}
\end{equation}
\begin{equation}
 B=H^{'}_{fs} \left(\sum_j\sum_k c_{jk} z_{sj}t_{fk}\sum_n v_{kn}\hat{X}^{-1}_{fn}X_{fn}\hat{X}^{-1}_{fn} \right)H^{'}_{fs}
\end{equation}
where $H^{'}_{fs}$ is the $H_{fs}$ of the previous update. To avoid scale ambiguity these normalizations should follow: $H_{fs} \leftarrow H_{fs}/tr(H_{fs})$, $t_{fk} \leftarrow t_{fk} / \sum_{f^*} t_{f^*k}$, $z_{sj} \leftarrow z_{sj}/ \sum_{j^*} z_{sj^*}$ and $c_{jk} \leftarrow c_{jk}/ \sum_{j^*} c_{j^{*}k}$. In the test phase a basis vector is kept fixed to a dictionary. Therefore, $c_{jk}=1$ only if the $k^{th}$ basis vector belongs to the $j^{th}$ dictionary, otherwise $c_{jk}=0$. Using equation \ref{CUpdtest} and the normalization, one can see that the values for $c_{jk}$ are then fixed for the whole iterative process.

  The estimated ID for speaker $s$ is $j$ for which $z_{sj}$ is maximal.
  \begin{equation}
  \label{SIDestZ}
  \hat{ID}_s=\argmax_{j} z_{sj}
  \end{equation}
Through $H_{fs}$ and $z_{sj}$, the speaker recognition can be interpreted as assigning the $j^{th}$ dictionary, and thus the target speaker identity of the $j^{th}$ dictionary , to the position of the $s^{th}$ speaker in the test utterance.
  Notice that $S$, the amount of speakers in the test mixture, can be equal to or smaller than $J$, the amount of speakers in the library. Not all known speakers must appear in the test mixture. However, for the experiments in this paper $S=J$.

  \section{Experiments}
  \label{experiments}
  \subsection{Data, set-up and parameters}
  The CHiME corpus has been used to perform our experiments \cite{Christensen10thechime}. It contains 34 speakers with 500 spoken utterances per speaker and a vocabulary of 52 words. Each utterance is about 1.5 seconds long. A scenario in time domain is simulated with a microphone array of $I=2$ microphones and three randomly chosen speakers ($S=3$) on randomly chosen spatial positions that are at least 20$^{\circ}$ apart. The RIR for each speaker to the microphones is determined via its spatial angle to the microphone array and some mild reverberation (RT\textsubscript{60} = 280 ms) \cite{campbell2005matlab}. 
  The microphones are placed 15 cm apart and sample at 16kHz. A Generalized Cross Correlation with Phase Transform (GCC-PHAT) is used to estimate the Direction of Arrival (DOA) and to perform a source count \cite{knapp1976generalized}. Aside from speaker recognition and source separation, also information on the number of sources and their location will be known. It has already been shown that such a GCC-PHAT is useful to initialize the multichannel NMF \cite{mirzaei2014blind}. Here it will be used to set up the initial off-diagonal elements of $\mathbf{H}_{fs}$. 
  
  For mixtures of 1.5 seconds, the GCC-PHAT finds the correct number of sources in about 85\% of the cases. This goes to 100\% for longer mixtures. Since this paper is not about source counting, mixtures with a false estimate of the number of sources will not be included in the results of the experiments in this paper.   
A spectrogram for each microphone is calculated using an STFT with a window length of 64 ms and an overlap of 32 ms.  

During the training phase each person speaks $U_{tr}$ utterances, without moving. The library is learned according to section \ref{NMFforSS} in 1000 iterations. For each speaker $K_{s}=K$ basis vectors are assigned, giving a total of $K_{tot}=KS$ basis vectors. Random initializations are used for $\mathbf{T}$ and $\mathbf{V}$. Basis vectors are fixed in a dictionary, belonging to a speaker on a certain location, by setting the speaker-indicators $\mathbf{Z}$ as follows.
\begin{equation}
z_{sk}=
\begin{cases}
1 & \text{ if $k\in \kappa^s$}\\
0 & \text{ otherwise}
\end{cases}
\end{equation}
Through $z_{sk}$ and the spatial information in $H_{fs}$, a basis vector is then fixed to a speaker on a certain location.

In the test phase the same speakers as in the training phase are used $(J=S=3)$, but placed on different virtual locations. Each person speaks $U_{test}=1$ utterance, without moving. The algorithm in section \ref{JointApproach} is applied and 1000 iterations are used. The library is taken from the training phase, $\mathbf{V}$ is initialized randomly and $\mathbf{H}$ is initialized using GCC-PHAT. The estimated IDs are determined by using equation \ref{SIDestZ}. 50 such test mixtures are created and for each three speakers have to be estimated. This gives a total of 150 recognition tasks per training set. In total twenty independent training sets are created
and their recognition accuracies are averaged to cope with the training variability. 

Pure SR performance is analyzed in the single speaker scenario, where no SS takes place and dictionaries are learned from non-reverberated speech. The \textit{joint} approach (section \ref{JointApproach}), which can be applied during either or both, testing and training, is compared with a single speaker scenario and a sequential scenario. In the \textit{single speaker} scenario, the dictionaries are trained or tested on non-reverberated speech of a single speaker, without applying SS. In the \textit{sequential} approach, SS is first applied on speaker mixture data, then dictionaries are trained or tested on the segregated streams.
In the remainder of the paper, the scenario where both training and testing are performed on simultaneous speech, will be referred to as the \textit{full simultaneous} scenario. In the \textit{full single} scenario both training and testing are performed on a single speaker.

\subsection{i-vector baseline}
A second baseline for a \textit{sequential} approach is also presented which uses i-vectors for the speaker recognition part. The MSR Toolbox is used to set up the speaker recognizer \cite{MSRToolbox}. 
First a Universal Background Model (UBM) and the total variability space are determined using 25 speakers not used in training phase. For every universal background speaker 500 single speech utterances are used. 
Source separation was performed on mixtures using the multichannel NMF with dictionary size $K=10$. The STFT features of the separated signals are transformed to MFCC features with differentials (MFCC$\Delta$) and accelerations (MFCC$\Delta\Delta$). i-Vectors are determined from the separated signals, the total variability space and the UBM. A test segment is classified using the cosine distance between the test i-vector and the training i-vectors. Table \ref{ivecrec} shows how the speaker recognition accuracies in the \textit{full simultaneous} scenario depend on the amount of UBM components $N_{comp}$ and the dimensionality of the i-vector $D_{ivec}$. The optimal values were found to be $N_{comp}=1024$  and $D_{ivec}=15$. This is comparable to the values found in \cite{drgas2015speaker}.  
\begin{table}
\centering
  \begin{tabular}{c|c c c c c c c}
    \multirow{2}{*}{$N_{comp}$} & \multicolumn{7}{c}{$D_{ivec}$} \\
  \hhline{~-------}
   & 5 & 10 & 15 & 20 & 50 & 100\\ 
   \hline
 256 & 68.3 & 73.7 & 74.6 & 74.5 & 62.5 & 62.4\\ 
   \hline
 512 & 66.1 & 74.8 & 74.8 & 73.4 & 68.8 & 66.5\\ 
   \hline
 1024 & 60.9 & 75.2 & \textbf{75.4} & 75.2 & 71.9 & 68.4\\ 
    \hline
 2048 & 55.4 & 68.5 & 70.6 & 72.7 & 73.8 & 72.2\\ 
  \end{tabular}
    \caption{\label{ivecrec}{i-vector based SR accuracies (in \%) for the \textit{full simultaneous} scenario.}}
  \end{table}
  \subsection{Results}



\begin{figure}
\centering
\includegraphics[width=0.5\textwidth]{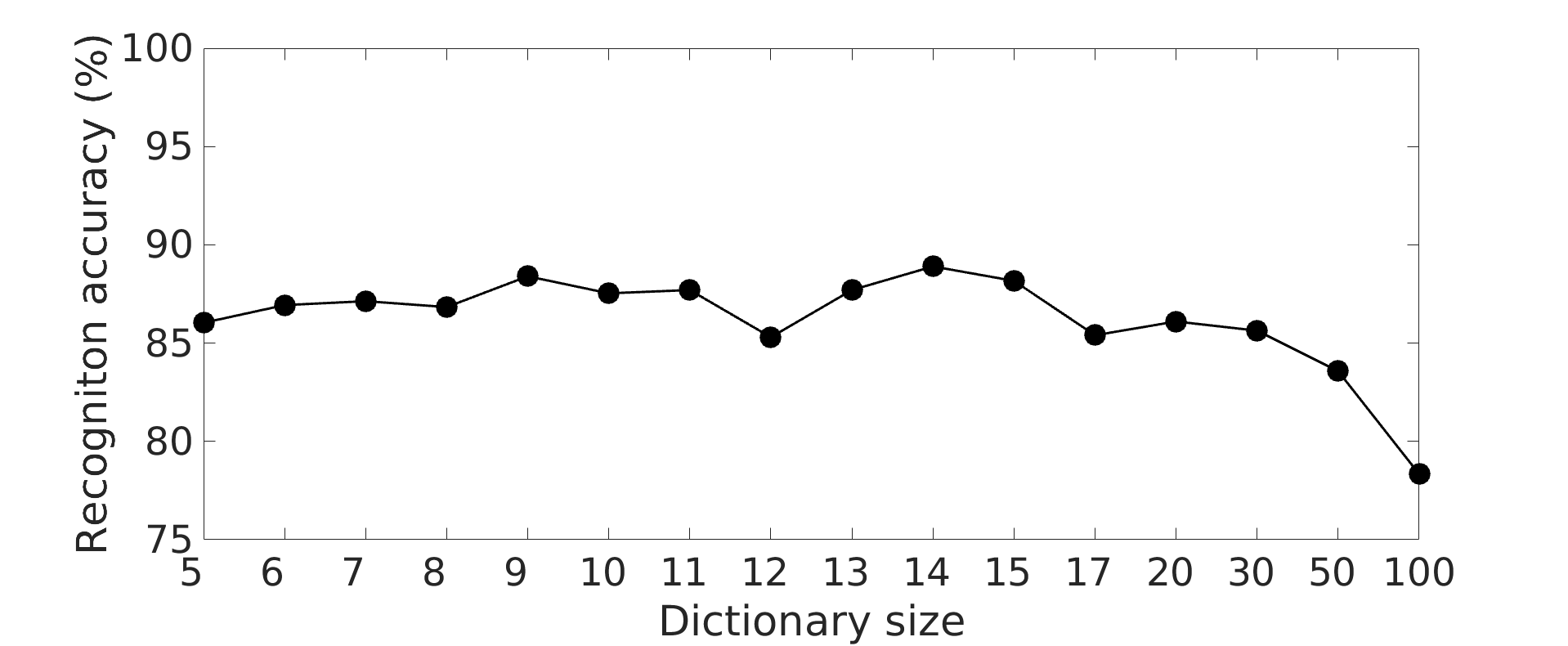}
\caption{\label{Kvarfig}{Average speaker recognition accuracy (in \%), depending on the dictionary size, using the \textit{joint} approach in the \textit{full simultaneous} scenario.}}
\end{figure}

In figure \ref{Kvarfig} speaker recognition accuracies relative to the dictionary size $K$ are plotted for the \textit{full simultaneous} scenario. The training size $U_{tr}$ was taken at 20 utterances per speaker. Results were obtained using the \textit{joint} solution. There is little influence of the dictionary size to the recognition accuracy if $K$ is between 8 and 30. In the remainder of the paper the dictionary size is taken at $K=10$. 

Speaker recognition accuracies for the different scenarios and used methods, using the above found optimal parameters ($K=10$, $N_{comp}=1024$ and $D_{ivec}=15$), are shown in table \ref{ResSR}(a) for NMF-based speaker recognition and table \ref{ResSR}(b) for i-vector based speaker recognition. 
In the \textit{full single} scenario i-vector based SR performs slightly better then NMF based SR. However, in \textit{simultaneous} scenarios, i-vectors struggle with the crosstalk after SS. The same problem occurs for NMF when a sequential approach is used in, either or both, training and testing. This problem is circumvented when the \textit{joint} approach for NMF in the \textit{full simultaneous} scenario is used.
The Speaker Error Rate (SER) for the \textit{joint} approach (12\%) is significantly lower than a \textit{sequential} approach using i-vectors for the SR (24\%). 

  \begin{table}
\centering
   \subfloat[NMF based SR.]{
\centering
\setlength\tabcolsep{4pt}
  \begin{tabular}{c|c|cc}
  \multirow{2}{*}{Train} & \multicolumn{3}{c}{Test} \\
  \hhline{~---}
   &  	single & seq & joint \\ 
 \hline
  single & 98.0  & \textbf{86.4} & 73.2 \\
  \hline
    seq & 88.8  & 76.1 & 62.5  \\
    joint & 80.2  & 61.3 & \textbf{87.9}\\
  \end{tabular}
  \label{ResNMFSR}
}\qquad
\subfloat[i-vector based SR.]{
\centering
\setlength\tabcolsep{4pt}
  \begin{tabular}{c|c|c}
    \multirow{2}{*}{Train} & \multicolumn{2}{c}{Test} \\
  \hhline{~--}
 & single & seq\\ 
 \hline
  single & \textbf{98.4}  & 78.5 \\
  \hline
    seq & \textbf{91.2}  & 76.0 \\
  \end{tabular}
  \label{ResIvecSR}
}

\caption{\label{ResSR}{SR accuracies (in \%) for different training and testing scenarios. In \textit{seq} and \textit{joint} 3 speakers are active simultaneously. \textit{Single} uses data from one speaker at a time.}}
\end{table}

\begin{figure}
\centering
\includegraphics[width=0.5\textwidth]{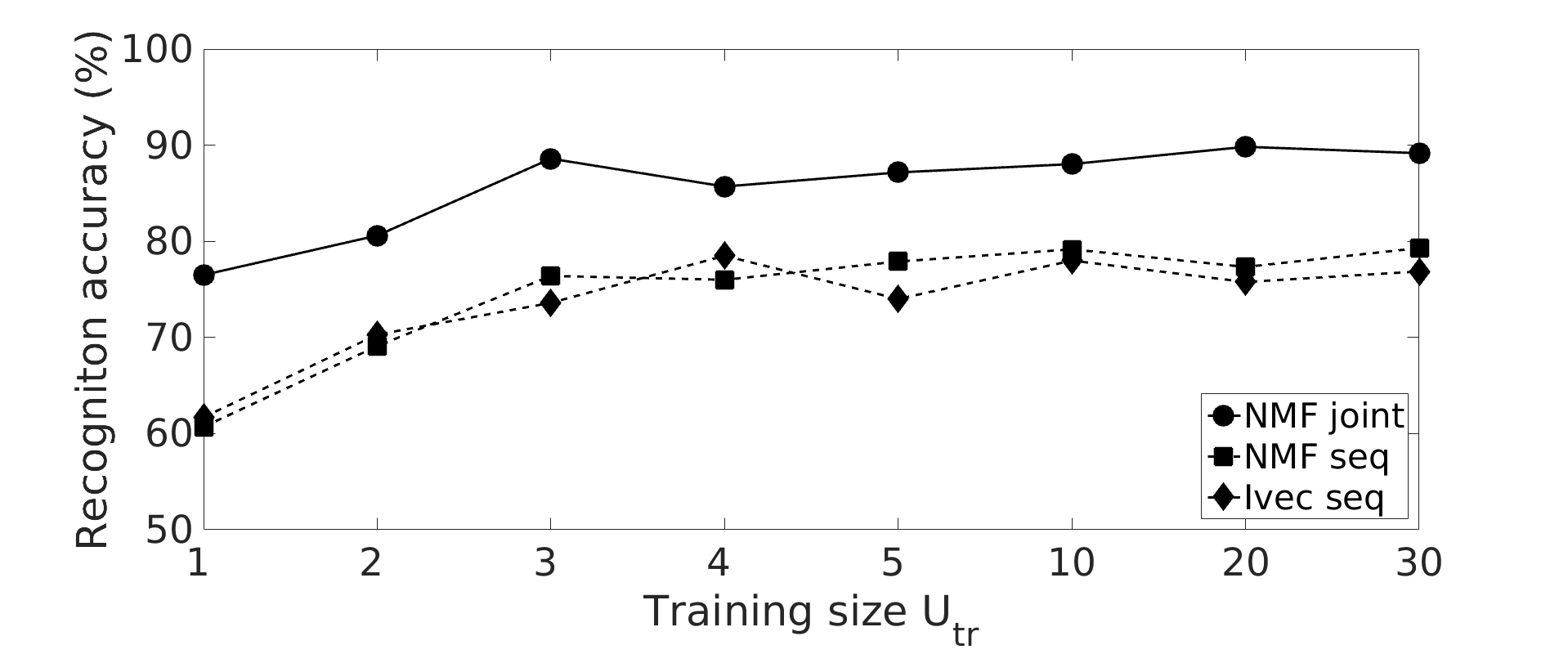}
\caption{\label{TrDatafig}{Average speaker recognition accuracy, depending on the training size $U_{tr}$, in the \textit{full simultaneous case}. Results are shown for the \textit{joint} approach (circles) and for a \textit{sequential} approach using NMF (squares) or i-vectors (diamonds).}}
\end{figure}

  Figure \ref{TrDatafig} shows how the recognition accuracy increases with the amount of training data. The optimal parameters above are used again. Only for very low amount of training utterances, the recognition accuracy decreases. The sufficiency of a limited amount of training data is probably due to the limited vocabulary size in the CHiME corpus. Notice that increasing the amount of training is beneficial twice to the recognition accuracy. SS is improved (see table \ref{SNR}) and the amount of patterns seen in training data is increased which allows the speakers to be better characterized and more easily recognized. To cope better with the increasing amount of seen patterns, the dictionary size could also be increased. This way the extra detected patterns can be stored in the dictionary. A system is chosen that can cope with any training size and thus the dictionary size is fixed.
  
  As this paper analyzes joint SR and SS, an evaluation metric for the SS quality is considered. Signal-to-distortion ratio (SDR), signal-to-interference ratio (SIR) and signal-to-artifact ratio (SAR) between the original source signal and the signal separated from the mixture, are calculated and are shown in table \ref{SNR} \cite{vincent2006performance}.
  Three different situations are shown: a mixture where each person speaks 20 utterances, a mixture where each person speaks 1 utterance and a mixture where each person speaks 1 utterance that is separated using previously learned dictionaries. 20 independent runs are used to calculate the ratios and the average is shown. As expected, the SS results are better when the mixture is longer. NMF has more examples to recognize recurring patterns and build distinctive dictionaries per speaker which enhances the separation quality. However, performing SS using learned and fixed dictionaries decreases performance. If a pattern is not seen in the training stage or is not frequent enough to fit in the dictionary, it cannot be used during testing for the reconstruction of the signal and will thus degrade the reconstructed signal. A solution is to allow flexibility in the trained dictionaries when testing, but this would interfere with the speaker recognition. 
    \begin{table}[ph]
\centering
\setlength\tabcolsep{3pt}
  \begin{tabular}{c|c c c}
 &  SDR & SIR & SAR\\ 
 \hline
  SS on 20 utts & 6.93  & 12.50 & 9.36 \\
  \hline
    SS on 1 utt & 4.28  & 7.99 & 8.26  \\
    \hline
    SS on 1 utt using learned dicts & 3.08 & 5.62 & 8.81 \\
  \end{tabular}
    \caption{SS performance for different scenarios measured in SDR, SIR and SAR (dB).}
    \label{SNR}
  \end{table}

  \section{Conclusion}
  \label{conclusion}
 In this paper it is shown that in simultaneous speech environments for three overlapping speakers, a joint approach for SS and SR outperforms sequential approaches for both NMF and i-vector based SR. This benefit is inherent to the multichannel NMF as patterns per speaker are learned anyway to perform a segregation. These patterns can be used to recognize speakers during the test phase.
 

  \newpage
  \eightpt
  \bibliographystyle{IEEEtran}

  \bibliography{mybib}
  \end{document}